# A particle can carry more than one bit of information

**Arindam Mitra**

Anushakti Abasan, Uttar-Phalguni -7, 1/AF, Salt lake,
Kolkata, West Bengal, 700064, India.

Abstract: It is believed that a particle cannot carry more than one bit of information. It is pointed out that particle or single-particle quantum state can carry more than one bit of information. It implies that minimum energy cost of transmitting a bit will be less than the accepted limit KTlog2.
.

## Introduction

In 1991, in a famous paper Bennett and Wisner pointed out [1] that due to entanglement two bits of information can be transmitted through an EPR particle, if the other particle of the EPR pair is sent in advance. The encoding is known as dense coding because *second* particle carries two bits of information due to prior transmission of the *first* particle of an EPR pair. Obviously, data transmission rate will be double during peak hours due to prior transmission of the first EPR particle during off-peak hours. The scheme has been partially experimentally verified [2].

Four distinguishable states are needed to represent four probable two-bit information : 00, 10, 01,11. As an EPR pair can be prepared exactly four distinguishable EPR states, Bennett and Wisner realized that an EPR pair can carry two bits of information.

$2^N$ distinguishable probable states are needed to represent N-bit message. One may think of using $2^N$ different single-particle states to transmit N–bit message. But the problem is, all the $2^N$ quantum states cannot be distinguished by the receiver due to no-cloning principle. Recently Brusβ et al pointed [3] out that Holevo bound shows that two qubits are necessary to transmit two-bit information. It appears that no-cloning principle and Holevo-bound restrict the transmission of more than one bit through a single quantum state. Next we shall see that no theoretical restriction can be given on the maximum number of bits transmitted by single particle or single quantum state.

In information processing time is an ignored dimension. But time can be directly utilized in information processing. Time gaps between successive particles can represent bit values. Four different time gaps between successive particles can represent two-bit information: $00 \to \nabla t_1$, $01 \to \nabla t_2$, $10 \to \nabla t_3$, $11 \to \nabla t_4$, where $\nabla t_1$, $\nabla t_2$, $\nabla t_3$ and $\nabla t_4$ are four probable time gaps between successive particles.

Sender has to transmit the first particle from a predetermined time. Receiver has to record the time of arrival of each particle. Receiver can recover the two-bit information from the time gaps between the successive arrival times of particles. Of course receiver has to recover the first two-bit information from the time gaps between the arrival time of the first particle and a predetermined time. Each particle carries two bits of information. Bit capacity C per particle is 2

All particles are assumed to be identical. So energy cost of transmitting each particle is same. Whatever be the minimum energy cost of transmitting a particle the minimum energy cost of transmitting a bit on the average will be less than the minimum energy cost of transmitting a particle. That is, with minimum energy cost nothing can be transmitted faster than information.

During the initial stage of development of classical information theory, it was pointed out [4-5] that there is a close relationship between thermodynamics and information theory. It is thought that at least KTlog2 energy is required to transmit a bit. Landauer first pointed out [6] that if bit is transported slowly by a vehicle, then energy requirement for bit-movement would be practically zero. He also noticed [7] that his idea didn't draw much attention of the scientific community. Actually Landauer considered the energy requirement of bit transportation, not bit transmission. Although Landauer's idea was bold, but it did not rule out the requirement of minimal energy cost of transmitting a bit of information. In the presented scheme single particle can carry more than one bit of information. So minimum energy cost of transmitting a bit of information will be less than the accepted limit KTlog2.

Suppose two single-particle quantum states $\psi$ and $\varphi$ represent two bit values. If $\psi$ and $\varphi$ are non-orthogonal states they cannot be distinguished with probability 1. Therefore, $\psi$ and $\varphi$ are supposed to be two orthogonal states. We know orthogonal quantum states can be distinguished with probability 1. But orthogonal quantum states cannot be distinguished with probability 1, if they arrive at different probable time gaps. It does not mean that time gaps cannot be encoded in orthogonal quantum states. We have to follow alternative quantum coding (AQC).

In AQC [8] two different sequence of quantum states represent two bit values. In AQC, time gaps between any two successive states in the two sequences are assumed to be same. Now we can choose different time gaps $\nabla t_i$ between two successive quantum states of the same sequence of two orthogonal quantum states comprising N states. Suppose we have two sequences, $S_a$ and $S_b$, of orthogonal states. Suppose, $2^m$ different time gaps $\nabla t_i$ are chosen for the sequence $S_a$ to represent multiple bits. Similarly, the same set of $2^m$ different time gaps $\nabla t_i$ are chosen for another sequence $S_b$ comprising N states to represent multiple bits. Total $2^{m+1}$ sequences each comprising N states, representing (m+1)-bit messages are probable. That is, each N-particle sequence represents each (m+1)-bit message. Bit capacity per particle C = (m+1)/N > 1 when (m+1) > N.

Let us illustrate the coding with the following example.

$$010... \longrightarrow S_a(\nabla t_1) = \{\psi(t_0 - t_1 = \nabla t_1), \quad \psi(t_2 - t_1 = \nabla t_1), \varphi(t_0 - t_1 = \nabla t_1), .........\}$$

$$110... \longrightarrow S_a(\nabla t_2) = \{\psi(t_0 - t_1 = \nabla t_2), \quad \psi(t_2 - t_1 = \nabla t_2), \varphi(t_3 - t_2 = \nabla t_2), ........\}$$

........................................................................................................................

$$001... \longrightarrow S_a(\nabla t_n) = \{\psi(t_0 - t_1 = \nabla t_n), \quad \psi(t_2 - t_1 = \nabla t_n), \varphi(t_3 - t_2 = \nabla t_n), ........\}$$

$$100... \longrightarrow S_b(\nabla t_1) = \{\varphi(t_0 - t_1 = \nabla t_1), \quad \psi(t_2 - t_1 = \nabla t_1), \psi(t_0 - t_1 = \nabla t_1), .........\}$$

$$111... \longrightarrow S_b(\nabla t_2) = \{\varphi(t_0 - t_1 = \nabla t_2), \quad \psi(t_2 - t_1 = \nabla t_2), \psi(t_3 - t_2 = \nabla t_2), ........\}$$

........................................................................................................................

$$000... \longrightarrow S_b(\nabla t_n) = \{\varphi(t_0 - t_1 = \nabla t_n), \quad \psi(t_2 - t_1 = \nabla t_n), \psi(t_3 - t_2 = \nabla t_n), ........\}$$

Suppose $\psi$ and $\varphi$ are $0°$ and $90°$ single-photon polarized states. Receiver knows the sequence codes in advance. Now receiver can identify $S_a(\nabla t_i)$ and $S_b(\nabla t_i)$ in the following way.

1. Receiver sets his analyzer either at $0°$ or $90°$ at fixed predetermined time. If he sets his analyzer at $0°$ then $90°$ photons will give null result, but $0°$ photons will give positive results. Receiver has to record the time of occurrence of positive results. Time of occurrence of null result cannot be known. Time difference between positive results will give time gaps. Minimum time gap between two positive results will be $\nabla t_i$.

2. Only from $\nabla t_i$ the sequence cannot be distinguished. Two different sequences $S_b$ and $S_b$ can have the same $\nabla t_i$. But all the time gaps between two positive results will not be same for the two different sequences $S_a(\nabla t_i)$ and $S_b(\nabla t_i)$, although minimum time gaps between two successive positive results will be same. The two different sequences of orthogonal states encoded in same time gaps can be identified by two different response timing.

References

1. C. H. Bennett and S. J. Wiesner, *Phys. Rev.Lett.* **96**, 2881 (1992).

2. K. Mattle et al , *Phys. Rev.Lett.* 76, 4656 (1996).

3. D. Bruβ et al. Int. Jour. of Q. info. July 18, 2:35 (2005) .

4. v. Neumann, *Theory of Self Reproducing Automata* ( Univ. Illiuonis Press, Urbana , 1960)

5. L. Brillouin, in *science and information Theory*, Chap. 13, p. 162,

6. R. Landauer , *Appl. Phy. Lett*, **51**, 2056 (1987);

7. R. Landauer, *Phy. Lett A*. **217**, 188 (1996).

8. A. Mitra, http://xxx.lanl.gov/cs.IT/0501023 , http://xxx.lanl.gov/cs.IT/0610102 ,
 http://xxx.lanl/cs.CR/0502026.